%% file: main.tex
\documentclass[10pt,conference]{IEEEtran}
\IEEEoverridecommandlockouts
\usepackage{cite}
\usepackage{amsmath,amssymb,amsfonts}
\usepackage{algorithmic}
\usepackage{graphicx}
\usepackage{textcomp}
\usepackage{xcolor}

\def\BibTeX{{\rm B\kern-.05em{\sc i\kern-.025em b}\kern-.08em
    T\kern-.1667em\lower.7ex\hbox{E}\kern-.125emX}}
    
\begin{document}

\title{Supporting Software Engineering Research and Education by Annotating Public Videos of Developers Programming}

\author{\IEEEauthorblockN{Abdulaziz Alaboudi}
\IEEEauthorblockA{\textit{George Mason University} \\
Fairfax, Virginia, USA \\
aalaboud@gmu.edu}
\and
\IEEEauthorblockN{Thomas D. LaToza}
\IEEEauthorblockA{\textit{George Mason University} \\
Fairfax, Virginia, USA \\
tlatoza@gmu.edu}
}

\maketitle
%
\begin{abstract}
Software engineering has long studied how software developers work, building a body of work which forms the foundation of many software engineering best practices, tools, and theories.  Recently, some developers have begun recording videos of themselves engaged in programming tasks contributing to open source projects, enabling them to share knowledge and socialize with other developers. We believe that these videos offer an important opportunity for both software engineering research and education.  In this paper, we discuss the potential use of these videos as well as open questions for how to best enable this envisioned use. We propose creating a central repository of programming videos, enabling analyzing and annotating videos to illustrate specific behaviors of interest such as asking and answering questions, employing strategies, and software engineering theories. Such a repository would offer an important new way in which both software engineering researchers and students can understand how software developers work.  
\end{abstract}

%
%

%
%

%

\input{sections/1.introduction}
\input{sections/2.examples}
\input{sections/3.work}

%
\section*{Acknowledgment}
This research was funded in part by NSF grant CCF-1703734.

%

%

\bibliographystyle{IEEEtran}
\bibliography{videoPaper}

\end{document}

%% file: sections/1.introduction.tex
\section{Introduction}
The artifacts developers generate as they work and coordinate with others have long offered an important window into developers' workflow, needs, and activities, offering an indirect means to observe developers through their committed code, issues, comments, social media posts, and other artifacts\cite{lakhani2004, Mamykina2011, singer2014Twitter, MacLeod2015}.

\begin{figure}
    \includegraphics[ width=\linewidth, keepaspectratio, clip, trim= 0cm 0cm 0cm 10cm]{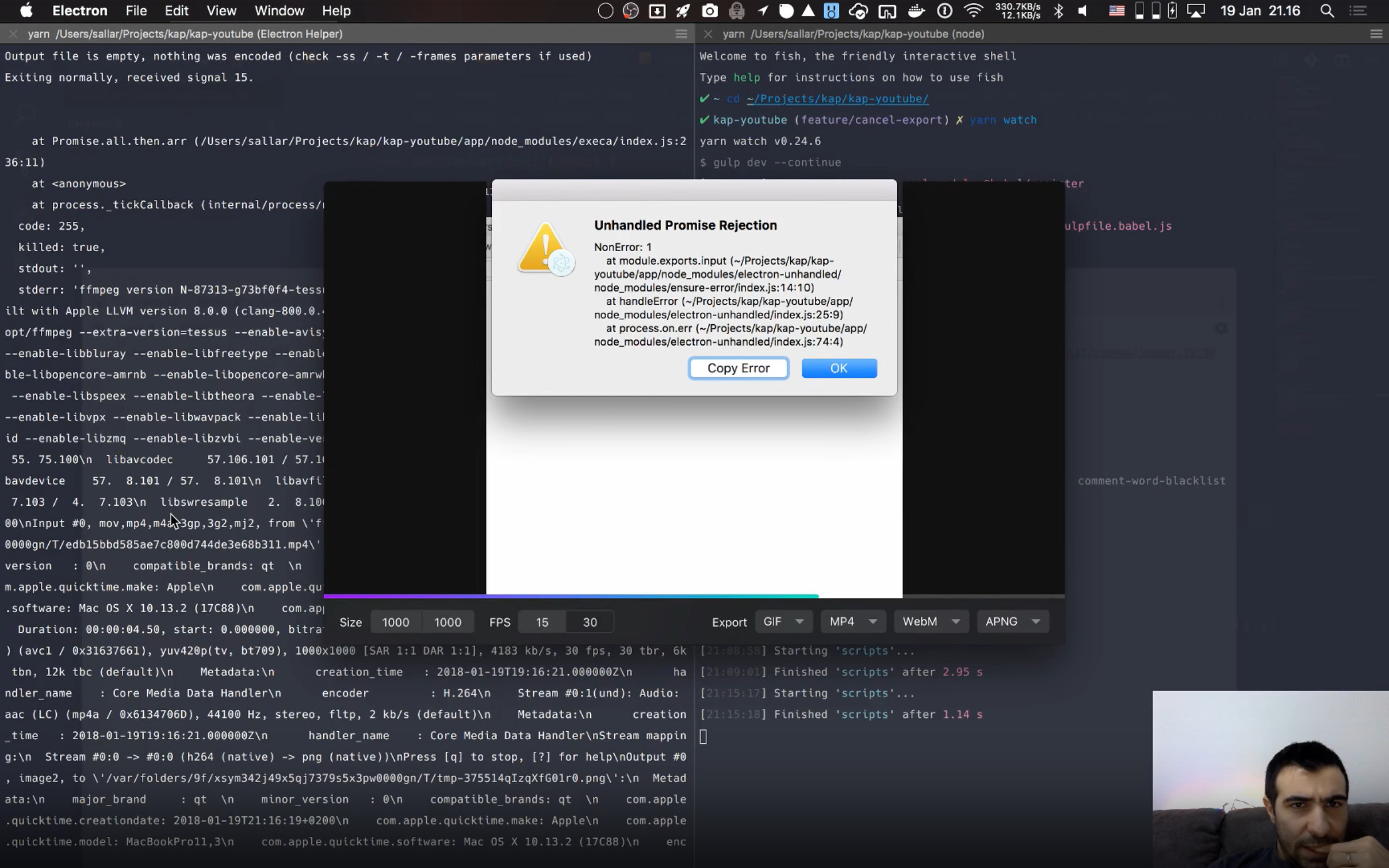}
    \caption{A video which shows a developer debugging.}
    \label{fig:example}
    \end{figure}

Recently, a new form of artifact has emerged: the screencast. Early use of screencasting was often intended to offer developers tutorial content, replacing traditional text-based documentation by explaining how to use development tools or new APIs \cite{ellmann2017find}\cite{MacLeod2015}.
More recently, developers have begun to live-stream their own real-time work on open source software\cite{Faas2018}. These videos illustrate developers' work in action using their preferred development environment while working on real tasks in familiar and unfamiliar code. These videos are not rehearsed and aim to shows a direct view of the moment-to-moment behavior of developers engaged in real software development work (Figure \ref{fig:example}).

By showing how developers work moment-to-moment,  these videos offer an essential resource for software engineering research and education \cite{Haaranen2017}. They enable direct observation of developers building, debugging, and testing software that would otherwise require conducting a field study. These videos may help to illustrate existing as well as potentially new software engineering theories, strategies, and best practices. To showcase how to use strategies for tasks such as debugging \cite{Irvin1989}, software engineering educators might make use of examples of developers at work in a real-world task rather than create artificial examples.

However, using public videos for research and education today is difficult. First, videos are scattered over the Internet. Developers have many options for hosting their videos such as YouTube and Twitch. As these are filled with millions of videos in other categories, finding programming videos is hard and reliant on the videos' titles. 
Second, to identify developers exhibiting a specific behavior in a context, activity, or strategy, one cannot search directly for such videos. Instead, it is necessary to watch videos at random, which are typically 1-6 hours long, and hope that the behavior of interest is exhibited by the developer.

Creating a central repository of programming videos for research and education offers a potential solution to these problems. Videos could be analyzed by the research community and annotated to reflect the behavior they contain, identifying specific contexts, techniques, issues, strategies, and theories that they illustrate. Researchers or educators might share specific lists of videos, and repository users could search for videos with specific characteristics.


%% file: sections/2.examples.tex
\section{MOTIVATING EXAMPLES}

To survey some of the potential benefits we envision of a central, annotated repository of programming videos, we describe several examples of its use for software engineering research and education. We name our proposed repository \textit{observe.dev}.

Sara is a professor who is teaching an undergraduate software testing class. She is planning to introduce test-driven development (TDD) to her students. She has prepared some materials for the class to teach the theory behind TDD. She also has made a simple example where she shows TDD in practice. However, she wants to also show how TDD is used in large projects and the practices, strategies, and tools developers use while applying it. She opens up \textit{observe.dev} and searches for videos of developers using TDD. The page lists several extended videos, including annotations for each denoting the time at which developers use TDD. She watches 20 minutes of video depicting developers practicing TDD and then shares this video with her students.

Deema is a new software engineering Ph.D. student who is trying to learn about theories of how developers navigate through code. While reading explanations of several theories in research papers, she discovers an information forging theory (IFT) paper which describes how developers navigate code while debugging \cite{Lawrance2013IFT}. She feels that her understanding of this theory is abstract, and a concrete example of a real developer navigating code that showcases this theory would help her more firmly grasp the concept. She uses \textit{observe.dev} to search for instances of developers browsing code using IFT. She finds 3 hours of a video illustrating a developer debugging within a large software project, with instances of IFT in action denoted with an annotation. After watching several minutes of a developer navigating code, she feels more confident in her understanding of this theory.      

George is a new Ph.D. student investigating the limitations of current debugging tools for web development and looking for a specific direction. Instead of trying to better understand how developers work with today's tools by conducting a lab study and recruiting participants to use a debugger, he decides to use \textit{observe.dev}. He searches for videos involving debugging tools for JavaScript using \textit{observe.dev} and finds over 50 instances. Watching these videos, he observes a number of instances in which the developer faces a similar challenge with these debugging tools. He annotates these instances and publishes them in \textit{observe.dev} to share them with the software engineering research community. Writing up a short paper on this observation, he suggests the need for a new form of debugging tool to address this challenge. He includes a link to videos with the new annotation for this challenge he added in \textit{observe.dev}.

%% file: sections/3.work.tex
\section{Preliminary work and Conclusion}
We have taken several initial steps towards creating a central repository of programming videos. We have collected over 40 hours of public programming videos. Our initial goal is to explore the value of these videos for research by investigating their use in understanding how developers debug and the strategies developers use which enable them to debug more effectively. 

In order to create a central repository of programming videos that serves both software engineering education and research, there are several important open questions which must be addressed.  What infrastructure and workflow is needed to effectively support and manage the contributions from the software engineering community? How can the annotated videos be effectively offered and displayed to students and instructors to facilitate their use in software engineering education? In what ways, if any, can tool support or automation make the process of curating and annotating videos which illustrate behaviors easier,  as some have begun to explore \cite{Ponzanelli2016, Moslehi2018}? Finally, what are the ethical implications of using these public videos of developers?  
We hope that a public repository of programming videos will provide a valuable resource for the software engineering community.